\documentclass[trackchanges,twocolumn]{aastex701}
\newcommand{\nuclei}[2]{\textbf{$^{#1}${#2}}}

\usepackage{float}
\usepackage{amsmath}
\begin{document}

\title{The Importance of Curium-250 to the $\gamma$-Ray Signature of $r$-Process Remnants}

\author[orcid=0000-0002-7893-4183]{Axel Gross}
\affiliation{GSI Helmholtzzentrum für Scherionenforschung, Planckstraße 1, D-64291 Darmstadt, Germany}
\affiliation{Theoretical Division, Los Alamos National Laboratory, Los Alamos, NM 87545, USA}
\email[show]{agross@lanl.gov}

\author[orcid=0000-0003-1758-8376]{Samuel Cupp}
\affiliation{X Theoretical Design, Los Alamos National Laboratory, Los Alamos, New Mexico 87545, USA}
\email{scupp@lanl.gov}  

\author[orcid=0000-0003-1758-8376]{G. Wendell Misch}
\affiliation{X Theoretical Design, Los Alamos National Laboratory, Los Alamos, New Mexico 87545, USA}
\email{wendell@lanl.gov}

\author[orcid=0000-0002-9950-9688]{Matthew R. Mumpower}
\affiliation{Obsidian Research, Fort Wayne, IN 46835, USA}
\affiliation{Department of Physics and Astronomy, University of Notre Dame, Notre Dame, IN, 46656, USA}
\affiliation{Computational and Artificial Intelligence Division, Los Alamos National Laboratory, Los Alamos, NM 87545, USA}
\affiliation{Center for Theoretical Astrophysics, Los Alamos National Laboratory, Los Alamos, NM 87545, USA}
\email{matthew@mumpower.net}

\begin{abstract}
Identifying the astrophysical sites of the heavy rapid neutron-capture process ($r$-process) remains one of the most prominent unsolved problems in nuclear astrophysics. 
Emitted $\gamma$-rays from the decay of unstable nuclei have been identified as a potential observable marker on both prompt and delayed timescales which could provide conclusive evidence of the presence of $r$-process activity. 
In this work, we identify and discuss the importance of \nuclei{250}{Cm} to the $\gamma$-ray spectra of remnants, showing that the decay spectrum of \nuclei{250}{Cm} dominates the $\gamma$-ray spectra of $r$-process ejecta across a broad timescale ($\sim$ 2,000 - 20,000 years). 
We highlight the uncertain nature of this spectra: \nuclei{250}{Cm} may decay via $\alpha$-decay, $\beta$-decay, and spontaneous fission, but the decay branchings are not experimentally known, and each decay channel produces unique spectral lines which may be observable.
We examine the potential for observation of the most prominent spectral signatures, finding that for conventional neutron star merger remnants, $\gamma$-ray lines will be detectable in current and future detectors only in the very near galaxy ($\sim$ 1 kpc), but other sites which could produce a larger ejecta mass may have greater observable distances. Our findings highlight the potential of actinide decay chains to indicate r-process nucleosynthesis and further underscore the necessity of measuring the nuclear properties that define their characteristic spectral signatures.
\end{abstract}


% =============================================
\section{Introduction} 
\label{sec:intro}
It is well-known that a significant fraction of the elements heavier than iron were synthesized in the rapid neutron capture process ($r$-process) (see e.g., \cite{1957RvMP...29..547B} , \cite{2026EPJA...62..105T}). 
Because the $r$-process occurs in environments with extreme conditions and involves capture on short-lived nuclei far from stability, it is difficult to obtain precise predictions for $r$-process production, and as a consequence, the astrophysical site(s) of the $r$-process have not been conclusively identified. 

The most studied sites of the $r$-process include Core-Collapse Supernovae ( \citealt{1994ApJ...433..229W,1996ApJ...471..331Q,2001ApJ...554..578W,2006ApJ...646L.131F}), and compact object mergers (see \citealt{1974ApJ...192L.145L,1989Natur.340..126E,1999A&A...341..499R,1999ApJ...525L.121F,2011ApJ...738L..32G,2012MNRAS.426.1940K,PhysRevD.87.024001,2013ApJ...773...78B,2014ApJ...789L..39W}); however, there are many proposed secondary sites, including magneto-rotational supernovae (\citealt{2012ApJ...750L..22W,2015ApJ...810..109N,2021Natur.595..223Y}), collapsars/hypernovae (\citealt{2013ApJ...778....8B,2019Natur.569..241S,2023PhRvD.108l3038A,2024ApJ...962...68A,MUMPOWER2025101736}), and magnetars (\citealt{2024MNRAS.528.5323C,2025ApJ...985..234P}). In particular, the site of the heavy $r$-process remains particularly uncertain, as the conditions necessary for a heavy $r$-process have not fully manifested in simulations of CCSNe \citep{2024ApJ...964L..16B,2025ARNPS..75..425J}, magneto-rotational supernovae \citep{2021MNRAS.501.5733R,2023MNRAS.518.1557R,2024MNRAS.529.3197R}, or magnetars \citep{2025ApJ...994...55P}.

Due to the challenges of understanding the $r$-process, there has been an increased focus on the potential for observational signatures to provide constraints on $r$-process sites. The landmark multi-messenger observations 
of GW170817 (see \citep{2017ApJ...848L..12A}) have provided evidence that NSMs are a site of the $r$-process; in particular, the observed kilonova, a radioactive transient thought to be powered by the decay of the synthesized, unstable $r$-process nuclei. The observed peak in the near-infrared and optical bands has been widely interpreted as evidence that NSMs undergo a strong $r$-process (\citealt{2017Sci...358.1570D,2017Natur.551...80K,2017Sci...358.1559K,2017ApJ...848L..34M,2017PASJ...69..102T,2018MNRAS.481.3423W,Brethauer:2024zxg}) due to the high opacities of ionized lanthanides (see e.g.,  \citealt{2022ApJ...939....8D,2025arXiv250707785F}), which when present, have been shown to result in a kilonova that peaks in the infrared (\citep{2013ApJ...774...25K,2013ApJ...775..113T}). In contrast, a weaker $r$-process has been shown to peak in the blue and optical band (\citealt{2010MNRAS.406.2650M,2011ApJ...736L..21R,2014MNRAS.441.3444M}). 

While we do not wish to understate the importance of kilonova observations and their significance in indicating the presence of an $r$-process at these sites, the multiphysics required in kilonova modeling presently make it challenging to make definitive conclusions about the nature of the $r$-process. While the widely argued interpretation that a red kilonova provides evidence of lanthanide production has merit, there are also alternative interpretations. For example, the observations of GRB211211A
and GRB230307 have also been shown to be consistent with a collapsar scenario which produces only a weak $r$-process component, despite an observed peak in the infrared \citep{2026ApJ..1002L..24R}.
There are additionally numerous poorly constrained model parameters which have been shown to have significant impact on the resultant light curve (see e.g., \citealt{2016ApJ...829..110B,2021ApJ...906...94Z,2021ApJ...910..116K,Fryer:2023pew,2023ApJ...944..144L}), which may change the interpretation of observable results. In particular, we highlight \cite{2023ApJ...958..121T} and \cite{Fryer:2023osz}, which showed that different ejecta velocity assumptions may lead to late-time features which mimic the presence of lanthanides. As a consequence of the uncertainties present, it may be challenging to conclusively identify the site of the heavy $r$-process from kilonova observations alone.

However, other direct observables of $r$-process nucleosynthesis have been suggested. In particular, as the ejecta becomes transparent in the days and weeks after the event, specific strong $\gamma$-ray lines from the decay of unstable nuclei may be directly observable (e.g.  \cite{1998ApJ...506..868Q,2016MNRAS.459...35H,2020ApJ...903L...3W,2020ApJ...889..168K,2022ApJ...933..111T,Vassh2024,2026ApJ..1004...93G}). The observation and identification of a known $\gamma$-ray decay line of an actinide would provide conclusive, direct evidence of heavy $r$-process activity, bypassing the difficult multiphysics required to understand kilonova light curves. This approach has its own challenges -- at prompt timescales, the ejecta is expanding rapidly, which leads to Doppler broadening of the spectral emission lines. This, combined with the large number of potential $\gamma$-ray lines at early timescales, makes it challenging to tie observed lines to a specific source nucleus \citep{2026ApJ..1004...93G}. It has also been suggested that $\gamma$-rays might be observable in remnant objects  when the ejecta has slowed and the number of nuclei decaying are minimized, which decreases the challenges of the identification of spectral lines, although, in this case, the incident flux is smaller and the observational range reduced \citep{2020ApJ...889..168K,2025ApJ...991..216A}.

In this work, we study and highlight the importance of \nuclei{250}{Cm} to the $\gamma$-ray spectra of remnants which are thousands of years old. \nuclei{250}{Cm} may undergo $\alpha$-decay, $\beta$-decay, or fission, but the decay branchings are not precisely known. Each of these decay channels produces unique spectral lines which might be observable in current and future $\gamma$-ray detectors, particularly if these spectral lines are more prominent than estimated. Although \nuclei{250}{Cm} has been identified as a potential source of $\gamma$-ray lines in other studies of remnants (\citealt{2020ApJ...889..168K,2025ApJ...991..216A}), the uncertainty of its spectral emission profile and its importance to the spectral emission has not been elucidated. In addition, one of the most prominent $\gamma$-ray lines has been incorrectly identified in past work, as the $\alpha$-decay chain of \nuclei{250}{Cm} populates the nuclear isomer \nuclei{246m}{Am} rather than the ground state, which dominantly emits a 1079 keV line instead of the reported 679 keV line. The observation of a $\gamma$-ray line resultant from \nuclei{250}{Cm} decay would be consequential, as it would represent direct detection of a heavy $r$-process.

We organize our paper as follows: In \S\ref{sec:decay}, we summarize the available \nuclei{250}{Cm} decay information. In \S\ref{sec:production}, we show that \nuclei{250}{Cm} is robustly produced in significant quantities in heavy $r$-process ejecta. In \S\ref{sec:spectra}, we examine the spectra of these ejecta, demonstrating the dominance of \nuclei{250}{Cm} in these spectra from $\sim$ 2000 - 20,000 years. In \S\ref{sec:observe}, we consider the potential for observability of different potential spectral lines resulting from \nuclei{250}{Cm} decay. We summarize and conclude in \S\ref{sec:conclude}.
% =============================================
\section{Curium-250 Decay Information}
\label{sec:decay}

The lifetime of \nuclei{250}{Cm} is known to be thousands of years, but precise decay information is uncertain. Decay is possible through $\alpha$-decay, $\beta^{-}$-decay, and spontaneous fission, and of those, only spontaneous fission has been observed. The partial fission half-life has been measured at $1.13 \pm 0.05 \times 10^4$ yr \citep{METTA19672822}, and also by others as $1.74 \pm 0.24 \times 10^4$ yr \citep{PhysRev.148.1192} and $\sim 2 \times 10^4$ yr \citep{PhysRev.107.1087}. The measurement of \cite{PhysRev.148.1192} was made using the same source as \cite{METTA19672822}, albeit with greater uncertainty in the mass spectroscopy to determine the \nuclei{250}{Cm} abundance, and \cite{PhysRev.107.1087} used extrapolation of other Cm isotopic abundances in the sample to determine the amount of \nuclei{250}{Cm} present. Therefore, \cite{METTA19672822} is considered the most reliable measurement.

Although the  $\beta^{-}$ and $\alpha$ half-lives have not been observed, they can be estimated. The small $Q_{\beta^{-}}$ of 38 keV requires that the $2^-$ ground state of \nuclei{250}{Bk} is populated directly, yielding a unique first-forbidden transition ($0^+ \rightarrow 2^-$) that requires $\log ft > 8.5$ \citep{2023ADNDT.15201584T} and consequently $\tau_{1/2}(\beta^-) \gtrapprox 0.67 \times 10^4$ yr. If $\log ft = 9.7$, as for \nuclei{250}{Bk} $\rightarrow$ \nuclei{250}{Cf}, then $\tau_{1/2}(\beta^-) \sim 10.6 \times 10^4$ yr. 
The $\alpha$-decay lifetime can also be estimated: from $r_0$ systematics (see e.g. \cite{AKOVALI19981}), $r_0 \sim 1.515$, which can be combined with $Q_\alpha = 5269$ keV, $I(g.s. \rightarrow g.s.) \sim 85\%$, and a hindrance factor of 1.0 to yield $\tau_{1/2}(\alpha) = 4.55 \times 10^4$ yr. The standard adoption of decay rates for \nuclei{250}{Cm} is $\tau_{1/2}\text{(SF)} = 1.13 \times 10^4$ yr, $\tau_{1/2}(\alpha) = 4.55 \times 10^4$ yr, and $\tau_{1/2}(\beta^-) = 10.6 \times 10^4$ yr, which yields a total half life of $\sim$ 8300 years with branching fractions 74\%, 18\%, and 8\% respectively.

\begin{figure*}[t]
\plotone{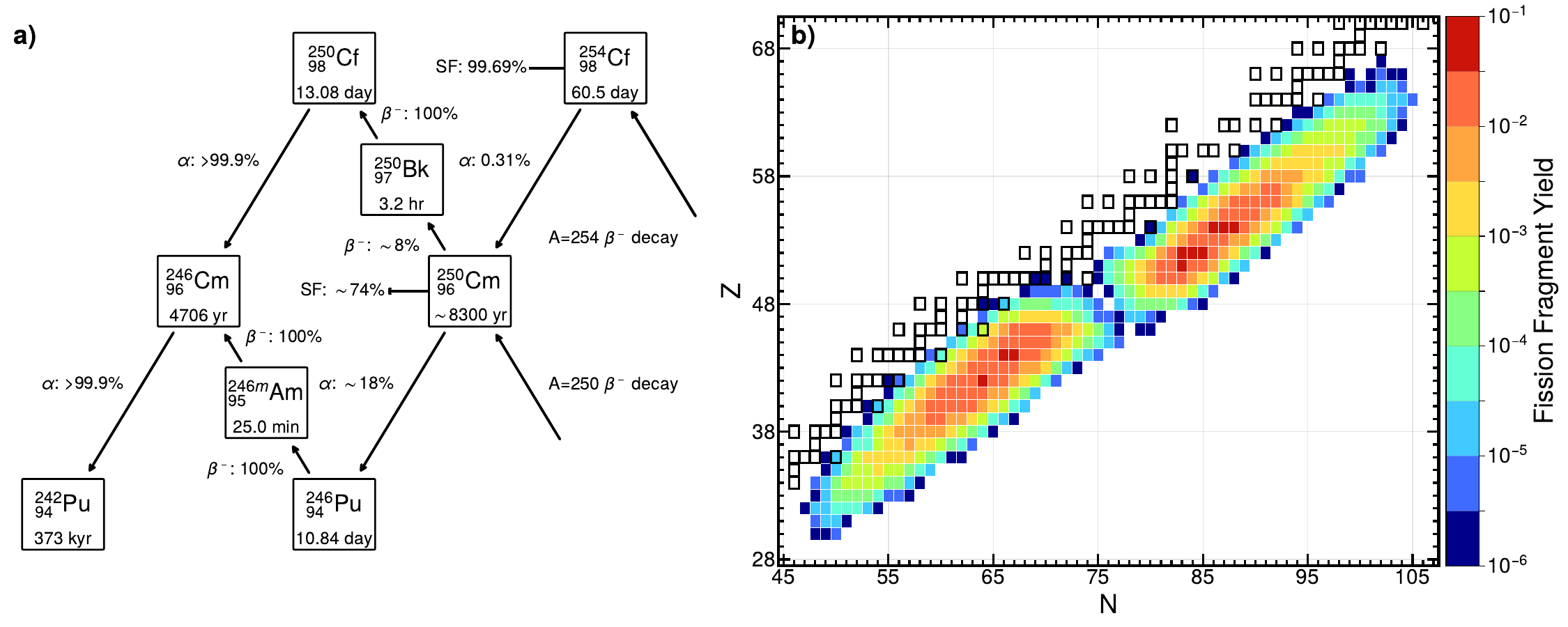}
\caption{\textbf{a)} Decay Pathways of \nuclei{250}{Cm}. Note that \nuclei{246m}{Am} is a $2^-$ nuclear isomer with energy $\sim 30$ keV, which directly $\beta$-decays to \nuclei{246}{Cm}, and that \nuclei{242}{Pu} undergoes $\alpha$-decay to \nuclei{238}{U}, which itself has a 4.5 Gyr half-life. \textbf{b)} Calculated fission fragment yield per unit decay of \nuclei{250}{Cm}. Black outlines denote stable nuclei.}
\label{fig:diagram}
\end{figure*}

We highlight the large uncertainties in the $\alpha$ and $\beta$ branchings which will have significant implications for $\gamma$-ray observables in the following sections, particularly if either of these branchings is larger than estimated. In Fig \ref{fig:diagram}a, we show the production and decay pathways of \nuclei{250}{Cm}. We note that the $\beta$-decay of \nuclei{246}{Pu}, along the $\alpha$-decay chain of \nuclei{250}{Cm}, produces the $\sim$ 30 keV isomeric state of \nuclei{246}{Am}, which itself directly undergoes $\beta$-decay and has a unique $\gamma$-ray spectra. This occurs because of the favorable transition between \nuclei{246}{Pu} and \nuclei{246m}{Am} ($0^+ \rightarrow 2^-$), as compared to the ground state ($0^+ \rightarrow 7^-$). In Fig \ref{fig:diagram}b, we show the fission yield of \nuclei{250}{Cm}. 
The fission fragment yields of \nuclei{250}{Cm} are calculated using the General Description of Fission Observables (GEF) code. 
GEF calculates the excitation energy partitioned to each primary fragment and models the subsequent prompt neutron and $\gamma$-ray emission via statistical de-excitation \citep{Schmidt2016}.

\section{Curium-250 Production in $r$-Process Ejecta}
\label{sec:production}

\nuclei{250}{Cm} is primarily produced astrophysically via direct $\beta$-decay of the A=250 chain, as the higher mass nuclei at $A=254$ and $A=258$ chain dominantly undergo spontaneous fission instead of $\alpha$-decay. As an exception, \nuclei{254}{Cf} $\alpha$-decays with a branching fraction of $0.31\%$, though this contributes an extremely small fraction of the produced \nuclei{250}{Cm}. As a consequence, the production of \nuclei{250}{Cm} is a direct diagnostic of the production of the $A=250$ $\beta$-decay chain. In addition, due to the short lifetime of the $\beta$-parent, \nuclei{250}{Am} ($\tau_{1/2}(\beta) = 72$ s), and its parents, \nuclei{250}{Cm} is promptly produced in the ejecta, and therefore the spectrum from its decay is determined entirely from its lifetime and decay properties. The decay of \nuclei{250}{Cm} can therefore be entirely decoupled from its production, and the characterization of observable $\gamma$-rays from \nuclei{250}{Cm} decay is entirely dependent on the total amount of \nuclei{250}{Cm} that is produced.

We find that significant amounts of \nuclei{250}{Cm} are produced across a wide variety of astrophysical conditions. To illustrate this, we perform a parametric study of conditions, with \nuclei{250}{Cm} set to be stable. We parameterize our trajectory in terms of the initial density of the ejecta, $\rho_0$, the initial electron fraction $Y_{e}$, and the expansion timescale, $\tau$.
The evolution of the density is parameterized by:
\begin{equation}
    \rho(t) = \begin{cases}
        \rho_0 e^{-t/\tau}, & t\leq t_{\rm tr}, \\
        \rho_0 (\frac{3\tau}{et})^3, & t>t_{\rm tr},
    \end{cases}
\end{equation}
where $t_{tr}=3\tau$, as suggested by \cite{lippuner2015}. 

We take an initial temperature of 9 GK, and assume the temperature evolves proportional to the cube root of the density:
\begin{equation}
\left(\frac{T(t)}{T_0}\right)^3 = \frac{\rho}{\rho_0}
\end{equation}
The initial composition is calculated based on the chosen initial electron fraction and density under the assumptions of nuclear statistical equilibrium using the SFHo equation of state \citep{2013ApJ...774...17S}.
We evaluate each parametric run using version 1.6.0 of the Portable Routines for Integrated nucleoSynthesis Modeling (PRISM) reaction network \citep{2021PhRvC.104a5803S}.  
The nuclear input to PRISM is based on the 2012 version of the Finite Range Droplet Model
\citep{2012PhRvL.108e2501M,2016ADNDT.109....1M}. 
Radiative capture and fission rates are calculated with the CoH$_3$ statistical Hauser–Feshbach code \citep{2019arXiv190105641K,2021EPJA...57...16K}. 
$\beta$-decay rates, including delayed neutron emission, are calculated assuming statistical de-excitation from excited states \citep{2016PhRvC..94f4317M,2018ApJ...869...14M}. 
The remaining reaction rates (e.g. alpha decay and other less important types) are obtained from the REACLIB database \citep{2010ApJS..189..240C}. Nuclear fission is handled as in \cite{Vassh2019}, except for \nuclei{250}{Cm}, which we implement as described above in \S\ref{sec:decay}. While the current version of PRISM does not support the inclusion of nuclear isomers, because we have identified that the $2^-$ state of \nuclei{246}{Am} is populated and relevant, we have manually modified the $\beta$-decay rate of \nuclei{246}{Am} to be the correct value, $\tau_{1/2} = 25.0$ min, and have similarly substituted the observed decay spectrum of \nuclei{246m}{Am} below.

We logarithmically sample $\rho_0$ between $10^6$ and $10^{9.5}$ $\text{g}/\text{cm}^3$, $\tau$ between $10^{-3.6}$ and $10^{-1}$ s, and uniformly sample $Y_{e}$ between 0.005 and 0.5. Each parameter takes 25 values (inclusive of both boundaries) for a total of 15,625 runs.

\begin{figure*}[t]
\plotone{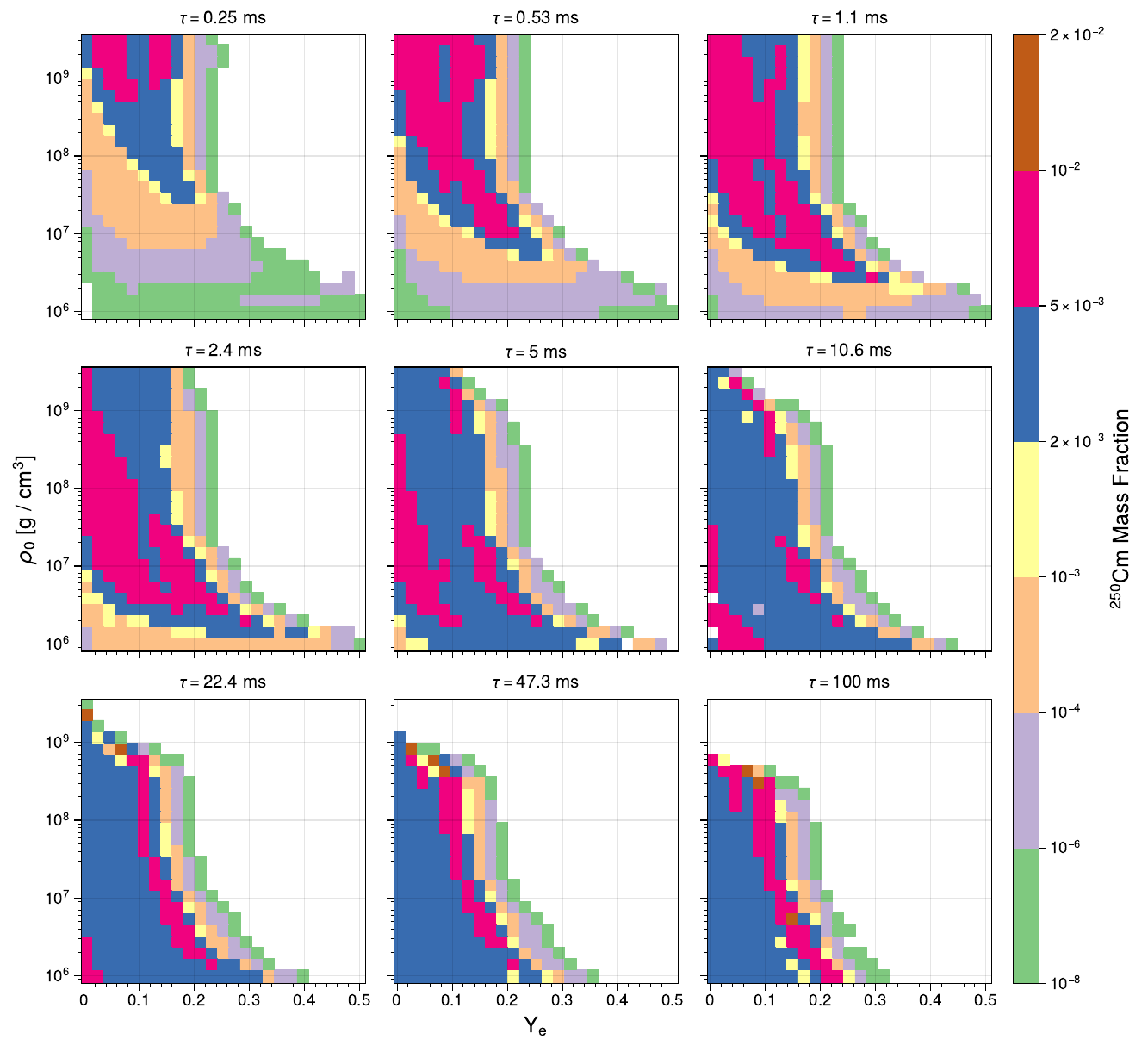}
\caption{Simulation results from parametric study of nucleosynthetic conditions in PRISM, showing the mass fraction of \nuclei{250}{Cm} as a function of the electron fraction ($Y_e$), initial density ($\rho_0$), and expansion timescale ($\tau$).}
\label{fig:parameter}
\end{figure*}

Fig. \ref{fig:parameter} shows the mass fraction of \nuclei{250}{Cm} as a function of $Y_e$, $\rho_0$, and $\tau$. It can be seen that large regions of the parameter space have a significant \nuclei{250}{Cm} mass fraction ($\gtrapprox$ 0.2\%). These regions correspond to the parameter space where a strong $r$-process occurs (see e.g. \cite{2025ApJ...990...37K}, and thus, \nuclei{250}{Cm} is robustly produced in significant quantities when the ejecta undergoes a strong $r$-process. In specific ejecta, we find the mass fraction to be as high as 2\%.

% =============================================
\section{Gamma-Ray Spectra}
\label{sec:spectra}
We calculate the $\gamma$-ray spectrum associated with $r$-process decays using the methodology of \cite{2026ApJ..1004...93G}, which calculates the emitted $\gamma$-ray spectra: $S(E,t)$, in units of number per unit energy per unit ejecta mass, as:

\begin{equation}
\label{eq:spectra}
S^{\gamma}(E,t)= \sum_{^A_ZX} F_{^A_ZX}(t) S^{\gamma}_{^A_ZX}(E),
\end{equation}
where $S^{\gamma}_{^A_ZX}(E)$ is the $\gamma$-ray emission of the nucleus ${^A_ZX}$ in number per unit energy per decay, and $F_{^A_ZX}(t)$ is the $\beta$-decay reaction flow as a function of time in decays per second per unit ejecta mass, and the sum is over every nucleus which decays in the network. The reaction flow is obtained from the PRISM network calculations. For the emission spectrum of individual nuclei, we use the $\gamma$-ray and x-ray spectra of ENDF/B-VIII.0 \citep{brown2018endfb}, if available. If there is no available $\beta$-decay spectrum for a specific nucleus, we use the tabulated results of \cite{MUMPOWER2025101736}, We also include the $\gamma$-ray and x-ray spectra for $\alpha$-decays if available in ENDF, and $\gamma$-rays from the prompt fissioning of nuclei as specified in \cite{2026ApJ..1004...93G}, except for \nuclei{250}{Cm}, for which we use the prompt fission spectrum as described above in \S\ref{sec:decay}.

\begin{figure*}[t]
\plotone{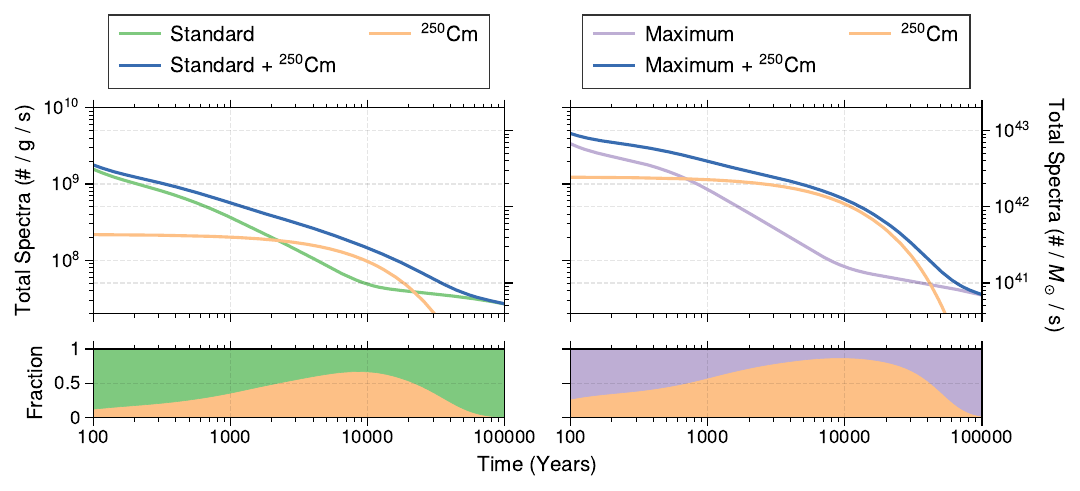}
\caption{Contributions to the $\gamma$-ray spectrum for two different ejecta: a ``standard" trajectory with a 0.3\% \nuclei{250}{Cm} mass fraction (left), and a ``maximum" trajectory with 2.0\% \nuclei{250}{Cm} mass fraction (right). Each panel shows the number of emitted $\gamma$-rays between 0.2 and 5 MeV as a function of time, and the fractional contributions from the decay pathways of \nuclei{250}{Cm} as compared to those from other nuclei.}
\label{fig:integrated}
\end{figure*}

We can modify this approach to include the separation of the \nuclei{250}{Cm} decays from the other decays of the ejecta:
\begin{equation}
S(E,t)= S_{ejecta}(E,t) + M_{frac}S_{Cm}(E,t),
\end{equation}
where $S_{ejecta}(E,t)$ is the spectrum calculated as in Eq.~\ref{eq:spectra} for the trajectory with stable \nuclei{250}{Cm}, $S_{Cm}(E,t)$ is the spectrum produced from the decay of 1 unit of \nuclei{250}{Cm}, and $M_{frac}$ is the mass fraction of \nuclei{250}{Cm} in the ejecta. In this manner, it is straightforward to quantify the $\gamma$-rays which are resultant from \nuclei{250}{Cm} decay and compare to those produced from the decay of other nuclei. 

We calculate the $\gamma$-ray spectrum for two of the trajectories from our parameterized study above: a “standard” strong $r$-process trajectory which produces a \nuclei{250}{Cm} mass fraction of 0.35\%, with parameters $Y_e = 0.11$, $\rho_0 = 2.05 \times 10^7$ $\textrm{g} \ \textrm{cm}^{-3}$, and $\tau = 10.6$ ms; and the trajectory which produces the maximum \nuclei{250}{Cm} mass fraction of 2.0\%, with parameters $Y_e = 0.07$, $\rho_0 = 4.22 \times 10^8$ $\textrm{g} \ \textrm{cm}^{-3}$, and $\tau = 100$ ms.

Fig. \ref{fig:integrated} displays the total number of $\gamma$-rays emitted between 0.2 and 5 MeV as a function of time for the two trajectories described above, as well as the fractional contribution to the emission from \nuclei{250}{Cm} and its decay chain. It can be seen that \nuclei{250}{Cm} and its decay chain dominate the production of $\gamma$-rays, producing over half of the emitted $\gamma$-rays in the ``standard" trajectory between  2200 and 22,000 years, peaking at 67\% of the emitted $\gamma$-rays. In the maximum trajectory, \nuclei{250}{Cm} produces over half of the $\gamma$-rays between  700 and  42,000 years, peaking at 87\% of the emitted $\gamma$-rays.
\begin{figure*}[t]
\plotone{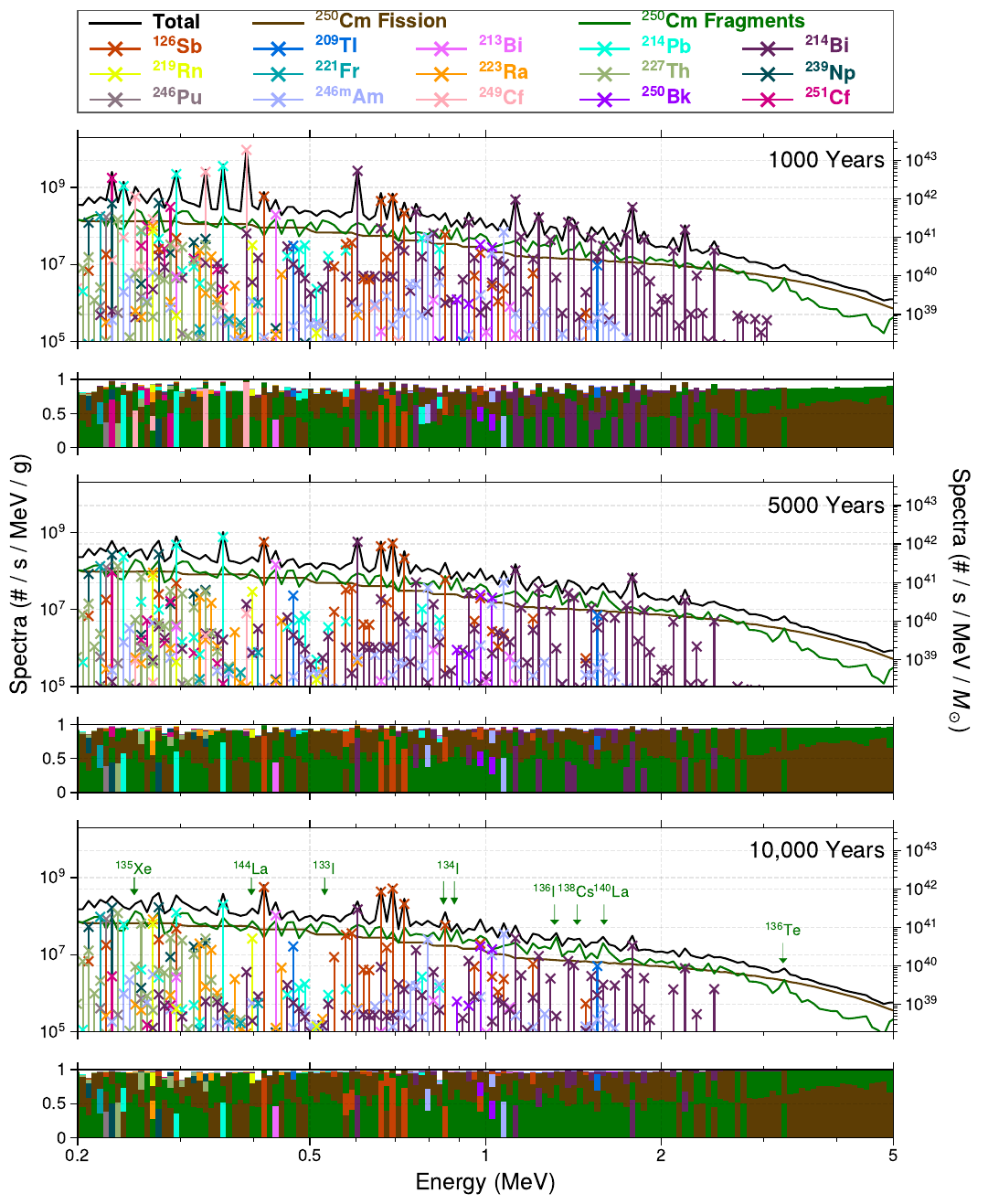}
\caption{$\gamma$-ray emission spectrum for the ``standard" trajectory at 1000, 5000, and 10,000 years. The solid black line gives the total spectra, while the colored crosses denote the contributions to this spectrum from an individual nucleus. The solid brown line gives the contribution from the prompt fission of \nuclei{250}{Cm}, while the dark green line gives the sum of contributions from the decay of unstable fission fragments produced by \nuclei{250}{Cm}. The source of the most prominent lines emitted from fission fragments are labeled in the lower panel.}
\label{fig:energyspectra}
\end{figure*}

In Fig. \ref{fig:energyspectra}, we show the energy-dependent $\gamma$-ray emission spectrum for the ``standard" trajectory at 1000, 5000, and 10,000 years, as well as the decomposition of this spectrum into the contributions from different nuclei. We summarize these contributions as follows:
\begin{itemize}
\item The prompt fission of \nuclei{250}{Cm} (brown line) produces approximately half of the ``background" which dominates the $\gamma$-ray emission.
\item The unstable fission fragments of \nuclei{250}{Cm} produce a forest of decay lines. The combination of these lines (dark green line) account for approximately half of the ``background" which dominates the $\gamma$-ray emission. There may be prominent lines from the decay of individual nuclei which are more significantly produced. We have marked the source of the most significant lines from fission fragments in the bottom panel of Fig. \ref{fig:energyspectra}. As the fission fragment distribution is uncertain, the exact prominence of these lines are unknown. 
\item The $\alpha$ and $\beta$ decay chains of \nuclei{250}{Cm} have a few significant $\gamma$-ray lines. From the $\beta$-decay chain, the decay of \nuclei{250}{Bk} produces a 989 keV line with 45\% intensity and 1032 keV line with 36\% intensity. From the $\alpha$-decay chain, the decay of \nuclei{246m}{Am} produces several lines, including a 1079 keV line at 28\% intensity, and the decay of \nuclei{246}{Pu} produces a 224 keV line at 24\% intensity. As the exact branching ratios for $\alpha$ and $\beta$ decay are uncertain, these lines could be significantly more prominent than shown here. 
\item Finally, there are a handful of lines resultant from the decay of other long-lived nuclei which are unconnected to \nuclei{250}{Cm}. The most notable of these are \nuclei{126}{Sb} (from \nuclei{126}{Sn}), \nuclei{214}{Bi} and \nuclei{214}{Pb} (from \nuclei{226}{Ra}), and \nuclei{249}{Cf} (from itself). We refer the reader to \cite{2026ApJ..1004...93G}, which provides a detailed accounting of the nuclei which emit $\gamma$-ray lines across broad timescales, including those shown here.
\end{itemize}
\begin{figure*}[t]
\plotone{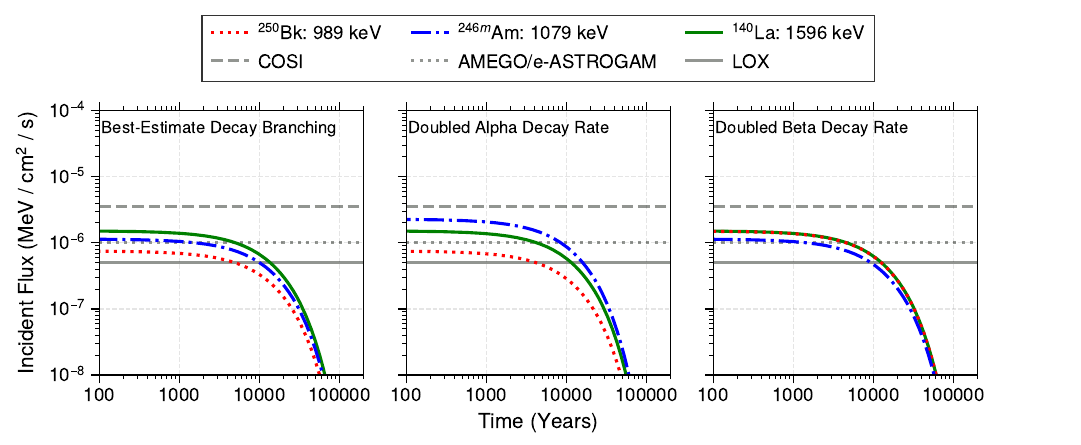}
\caption{Incident Flux of Prominent $\gamma$-ray lines for a \nuclei{250}{Cm} mass of 0.0002 $M_{\odot}$ at a distance of 1 kpc for three different scenarios: the ``best-estimate" decay branchings, when the $\alpha$-decay rate is doubled relative to this case, and when the $\beta$-decay rate is doubled relative to this case. Horizontal lines denote estimated detector sensitivities at 1 MeV for future and proposed $\gamma$-ray observatories: COSI \citep{tomsick2023comptonspectrometerimager}, AMEGO \citep{mcenery2019allskymediumenergygammaray}, e-ASTROGAM \citep{DEANGELIS20181}, and LOX \citep{miller2019exlunascientialunar}.}
\label{fig:observe}
\end{figure*}
Our findings establish \nuclei{250}{Cm} as the dominant source of emitted $\gamma$-rays from thousands to tens of thousands of years in ejecta which undergo a strong $r$-process. We find that all the emitted $\gamma$-rays--from prompt fission, fission fragments, alpha, and beta decays, are controlled by the \nuclei{250}{Cm} decay timescale, as the shorter-lived products maintain secular equilibrium with \nuclei{250}{Cm}. We therefore expect the $\gamma$-ray emission signal from these ejecta to be nearly constant from $\sim$ 2000 to $\sim$ 20,000 years, and almost entirely composed of the expected emission from \nuclei{250}{Cm}. 

The nature of this $\gamma$-ray emission signal is less certain. As mentioned above, the branching ratios for $\alpha$ and $\beta$ decay are estimated. If either of these ratios are different than expected, the corresponding $\gamma$-ray emission lines may be stronger or weaker. In addition, the fission fragment distribution is uncertain, which will affect the strength of the most prominent $\gamma$-ray lines from fragments. 

More generally, a remnant object may have ejecta which possess a variety of conditions, including weaker $r$-process components. In those ejecta, the only significant spectral lines at the relevant timescales come from the decay of \nuclei{126}{Sb}. This will have the effect of increasing the prominence of the \nuclei{126}{Sb} lines relative to the spectrum shown in Fig. ~\ref{fig:energyspectra}.

% =============================================
\section{Observability of Decay Lines}
\label{sec:observe}
In this section, we consider the potential for observation of strong $\gamma$-ray lines from \nuclei{250}{Cm} in remnants. As discussed above, the $\gamma$-ray signal of \nuclei{250}{Cm} can be decoupled from the specific details of the ejecta and expressed as a function of the total amount of \nuclei{250}{Cm} produced. For a specific $\gamma$-ray line, the incident flux can be expressed as:
\begin{equation}
\label{eq:IncidentFlux}
F_\gamma(t)=  \frac{M_{Cm}E_\gamma R_\gamma(t)}{4\pi D^2},
\end{equation}
where $M_{Cm}$ is the total \nuclei{250}{Cm} mass synthesized in the remnant, $E_\gamma$ is the energy of the line, $R_\gamma(t)$ is the time-dependent production rate of the $\gamma$-ray line per unit \nuclei{250}{Cm}, and $D$ is the distance to the remnant. 

We consider three of the most prominent potential lines in our analysis: the 989 keV line from the decay of \nuclei{250}{Bk} which is produced in the $\beta$-decay of \nuclei{250}{Cm}, the 1079 keV line from the decay of \nuclei{246m}{Am} which is produced in the $\alpha$-decay of \nuclei{250}{Cm}, and the 1596 keV line of \nuclei{140}{La}, which is produced as a fission fragment of \nuclei{250}{Cm}. Each of these lines comes from a different decay channel, and to illustrate potential variations in the expected flux from the uncertainty of the decay branchings, we consider two alternative decay branchings: where the $\alpha$-decay rate is doubled to give $\tau_{1/2}(\alpha)=2.275 \times 10^4$ yr, which yields a total half life of $\sim$ 7000 years and branching fractions 62\%, 31\%, and 7\% for fission, $\alpha$, and $\beta$, respectively; and where the $\beta$-decay rate is doubled to give $\tau_{1/2}(\beta^{-}) = 5.3 \times 10^4$ yr, which yields a total half life of $\sim$ 7700 years and branching fractions 68\%, 17\%, and 15\% for fission, $\alpha$, and $\beta$, respectively.

Fig. \ref{fig:observe} shows our results for nominal values of $M_{Cm} = 2 \times 10^{-4}$ $M_{\odot}$ and $D$ = 1 kpc, as compared to the expected sensitivities of upcoming and proposed $\gamma$-ray observatories. With an average \nuclei{250}{Cm} mass fraction between 0.002 and 0.01 in actinide-rich ejecta (see Fig. \ref{fig:parameter}), this corresponds to $\sim$ 0.02 - 0.1 $M_\odot$ of actinide-rich ejecta. It can be seen that even with a significant amount of produced actinides, observation would be a rare event. 

The most conventional candidate events for an actinide-rich $r$-process are NS-NS and BH-NS mergers, which may produce up to 0.1 $M_\odot$ of ejecta (e.g. \cite{2020ApJ...893..153K}). With an estimated occurrence rate in our galaxy of $\sim 10^{-4}-10^{-6}$ per year \citep{2026ApJ..1006L..49F}, it is unlikely that a suitable remnant exists in the observable region. However, other proposed $r$-process sites may produce higher ejecta mass. For example, it has been proposed that the cocoon of a $\gamma$-ray burst may produce neutrons sufficient for neutron-capture nucleosynthesis \citep{2025ApJ...982...81M}. The ejected mass of such events can be much more significant, which may substantially increase the observable range of $\gamma$-ray signals, especially if the $\alpha$ or $\beta$ decay branchings are larger than estimated, perhaps to the order of ($\sim$ 5 kpc). 

Practically, the most prominent spectral lines from a remnant of suitable age are those from the decay of \nuclei{126}{Sb}. While a detection of decay lines from \nuclei{126}{Sb} would provide evidence of the presence of $r$-process nucleosynthesis in the remnant, it would provide little information as to the nature of the $r$-process which occurred, as \nuclei{126}{Sb} is created in substantial quantities for any ejecta which reaches the 2nd peak of the $r$-process. In contrast to \nuclei{126}{Sb}, the detection of lines resultant from \nuclei{250}{Cm} would be extremely informative, as it would provide evidence that a strong $r$-process has occurred. For these reasons, we emphasize the importance of the accurate measurement of the lifetime and decay branchings of \nuclei{250}{Cm}, as it will clarify which $\gamma$-ray lines are expected to be most prominent and enable understanding of the nature of the ejecta if these lines are observed in remnants. These decay measurements will be possible in future experimental facilities such as the N=126 factory \citep{SAVARD2020258}, and due to their importance, we suggest that they are prioritized. 
% =============================================
\section{Conclusions}
\label{sec:conclude}
In this work, we have explored the impact of \nuclei{250}{Cm} on the $\gamma$-ray emission spectra of remnants. We find that \nuclei{250}{Cm} is consistently created in significant quantities for ejecta which produce a strong $r$-process, and that the spectrum resultant from \nuclei{250}{Cm} decay dominates the $\gamma$-ray emission from $\sim$ 2,000 - 20,000 years. Due to uncertain decay branchings, the nature of this spectrum is not precisely determined, but prominent lines from all decay channels may be observable in current and future $\gamma$-ray detectors at nearby distances ($\sim$ 1 kpc), or farther in the case of collapsars ($\sim$ 10 kpc). Detection of any of these lines would represent conclusive evidence of a heavy $r$-process, highlighting the importance of accurately characterizing these spectral signatures. As a compelling case study, \nuclei{250}{Cm} illustrates the broader potential of actinide observables across various timescales, underscoring the need for measurements of actinide and superheavy decay data at future experimental facilities.

\begin{acknowledgments}

\noindent LANL is operated by Triad National Security, LLC, for the National Nuclear Security Administration of U.S. Department of Energy (Contract No. 89233218CNA000001). 
Research presented in this article was supported in part by the Laboratory Directed Research and Development program of Los Alamos National Laboratory under project number 20240004DR.

\end{acknowledgments}

\bibliography{refs}{}
\bibliographystyle{aasjournalv7}
\end{document}